\documentclass[aps,pre,groupedaddress,showpacs]{revtex4} %
\usepackage{graphicx}
\begin{document}

\title{Polychromatic solitons in a quadratic medium}
\author{I. N. Towers and B. A. Malomed}

\affiliation{Department of Interdisciplinary Studies,
Faculty of Engineering, Tel Aviv University,
Tel Aviv, Israel}

\begin{abstract}
We introduce the simplest model to describe parametric
interactions in a quadratically nonlinear optical medium with the
fundamental harmonic containing two components with (slightly)
different carrier frequencies [which is a direct analog of
wavelength-division multiplexed (WDM) models, well known in media
with cubic nonlinearity]. The model takes a closed form with three
different second-harmonic components, and it is formulated in the
spatial domain. We demonstrate that the model supports both
polychromatic solitons (PCSs), with all the components present in
them, and two types of mutually orthogonal simple solitons, both
types being stable in a broad parametric region. An essential
peculiarity of PCS is that its power is much smaller than that of
a simple (usual) soliton (taken at the same values of control
parameters), which may be an advantage for experimental generation
of PCSs. Collisions between the orthogonal simple solitons are
simulated in detail, leading to the conclusion that the collisions
are strongly inelastic, converting the simple solitons into
polychromatic ones, and generating one or two additional PCSs. A
collision velocity at which the inelastic effects are strongest is
identified, and it is demonstrated that the collision may be used
as a basis to design a simple all-optical XOR logic gate.
\end{abstract}
 \maketitle

\section{Introduction}

Wave mixing at different carrier frequencies, of which generation of higher
harmonics is a well-known example, has had a long history of investigation
\cite{Franken61,Armstrong62}. In an optical medium whose symmetry group
lacks a center of inversion, the lowest-order nonlinear response is
quadratic, which gives rise to three-photon interactions. The resultant
three-wave mixing is resonant when the condition $\omega _{3}=\omega
_{1}+\omega _{2}$, imposed on frequencies of the three waves, is met. In the
particular case when $\omega _{1}=\omega _{2}$, the process reduces to the
second harmonic generation (SHG).

Solitons in quadratic [$\chi ^{(2)}$] media has been a major area of
research recently (see Ref. \cite{Etrich01} for a review). The first
experimental observation of $\chi ^{(2)}$ solitons was reported for type-I
SHG, which involves exactly one component of each harmonic, in the $(2+1)$%
-dimensional (bulk) geometry \cite{Torruellas95}. The observation
of solitons in the $(1+1)$-dimensional geometry (planar nonlinear
waveguide) followed shortly afterwards \cite{Schiek96}. Much
theoretical work has been performed for the solitons in both
type-I and type-II SHG, the latter case involving two components
of the fundamental harmonic, corresponding to different
polarizations, and a single component of the second harmonic \cite
{Stegeman,Buryak95,Tran95,Malomed96,Buryak97,JenaTypeII}. The SHG
process in an isotropic medium, with two polarizations at both
harmonics, has been considered too \cite{Boardman}.

More complex cases of multi-resonance wave mixing in quadratically nonlinear
media have not received much attention because of serious difficulties with
their experimental realization using the birefringence-based wave-vector
matching schemes. However, the recent rapid development of the
quasi-phase-matching (QPM) technique has changed the situation. The
technique, originally proposed long ago \cite{Armstrong62}, relied upon
periodic structures (usually, periodically poled ones) with an alternating
sign of the quadratic nonlinearity. Recently, the QPM technique has been
extended from periodic to Fibonacci-series-based structures \cite{Fradkin99}%
, and further to fully quasiperiodic ones \cite{Fradkin02} (see also Ref.
\cite{Johansen}). This makes it possible to essentially relax conditions on
the wave-vector difference for the matching to take place. Results presented
in Ref. \cite{Fradkin02} show, both theoretically and experimentally, not
only high values of effective $\chi ^{(2)}$ coefficients, but also that one
can attain the wave-vector matching simultaneously for several sets of waves
involved in the nonlinear interactions. In particular, the development of
the QPM technique has been an incentive to study multi-resonance systems
\cite{Kivshar99a,Kivshar99b,Towers00,Sukhorukov01}, a possible application
of which may be design of soliton-based logic gates. Indeed, a soliton may
naturally be used as a bit of information, and the interactions of solitons
can potentially support logic operations.

In this work, we develop a model of nonlinear mixing between two
fundamental-harmonic waves with different frequencies in a
quadratic medium. Via the $\chi ^{(2)}$ nonlinearity, they
generate three different wave components of the second harmonic.
Note that interactions between waves with different frequencies in
optical media with cubic ($\chi ^{(3)}$) nonlinearity is a
well-known topic, which has extremely important applications to
the WDM multichannel format of data transmission in fiber-optic
telecommunication links \cite{Agrawal}; however, it appears that
the issue has not yet been studied for $\chi ^{(2)}$ media, in
which the mixing may be realized in both spatial and temporal
domains.

We formulate the model in section \ref{two}, and produce its
stationary soliton solutions in section \ref{three}. These may be
both fully polychromatic solitons (PCSs), including all the five
field components, and special (simple) solutions of ``A'' and
``B'' types, which amount to ordinary two-component SHG solitons
in mutually orthogonal (non-intersecting) subsets of the five-wave
system. In section \ref{four}, we demonstrate, by means of direct
simulations, that the solitons of all these types are stable in a
broad region in the system's parameter space. In the same section,
we simulate collisions between the simple solitons. When they
overlap, the nonlinear interaction generates a component which was
absent in both of them before the collision, which makes the final
result of the collision strongly inelastic: the former simple
solitons develop extra components and become polychromatic ones.
Additionally, one or two extra solitons are generated by the
collision. We demonstrate that the collision between the initial
simple solitons may be a basis for an all-optical XOR logic gate.
The paper is concluded by section \ref{five}.

\section{The model}\label{two}

We consider the interaction of five waves in a diffractive dielectric medium
with the quadratic nonlinear susceptibility. The carrier frequencies of the
waves satisfy resonant conditions, $\omega _{3}=2\omega _{1},\;\omega
_{4}=\omega _{1}+\omega _{2},\;\omega _{5}=2\omega _{2}$, so that $\omega
_{1}$ and $\omega _{2}$ may be classified as carriers of the
fundamental-harmonic components, while $\omega _{3}$, $\omega _{4}$ and $%
\omega _{5}$ represent three components of the second-harmonic wave group,
generated by the two fundamental-harmonic components via the $\chi ^{(2)}$
nonlinearity.

Assuming, as usual, that the wave envelopes $E_{1},E_{2}$ and $%
E_{3},E_{4},E_{5}$ of these components are slowly varying ones, a system of
five equations, coupled parametrically through the components of the $\chi
^{(2)}$ nonlinear susceptibility tensor, can be derived from the Maxwell's
equations to govern the evolution of the waves in the spatial domain (the
derivation follows the well-known procedure worked out for the usual type-I
and type-II $\chi ^{(2)}$ interactions, see a detailed account given in the
review \cite{Etrich01}):

\begin{eqnarray}
2ik_{1}\frac{\partial E_{1}}{\partial z}+\frac{\partial ^{2}E_{1}}{\partial
x^{2}}+2\tilde{\chi}_{1}E_{1}^{\ast }E_{3}e^{i\Delta k_{3,1,1}}+\sqrt{2}%
\tilde{\chi}_{2}E_{2}^{\ast }E_{4}e^{i\Delta k_{412}} &=&0,  \nonumber \\
2ik_{2}\frac{\partial E_{2}}{\partial z}+\frac{\partial ^{2}E_{2}}{\partial
x^{2}}+\sqrt{2}\tilde{\chi}_{2}E_{1}^{\ast }E_{4}e^{i\Delta k_{412}}+2\tilde{%
\chi}_{3}E_{2}^{\ast }E_{5}e^{i\Delta k_{522}} &=&0,  \nonumber \\
2ik_{3}\frac{\partial E_{3}}{\partial z}+\frac{\partial ^{2}E_{3}}{\partial
x^{2}}+\tilde{\chi}_{1}E_{1}^{2}e^{-i\Delta k_{311}} &=&0,  \label{startsys}
\\
2ik_{4}\frac{\partial E_{4}}{\partial z}+\frac{\partial ^{2}E_{4}}{\partial
x^{2}}+\sqrt{2}\tilde{\chi}_{2}E_{1}E_{2}e^{-i\Delta k_{412}} &=&0,
\nonumber \\
2ik_{5}\frac{\partial E_{5}}{\partial z}+\frac{\partial ^{2}E_{5}}{\partial
x^{2}}+\tilde{\chi}_{3}E_{2}^{2}e^{-i\Delta k_{522}} &=&0,  \nonumber
\end{eqnarray}
where $k_{1}$ through $k_{5}$ are the corresponding carrier wave numbers, $z$
and $x$ are the propagation and transverse coordinates, $\Delta
k_{lmn}=k_{l}-k_{m}-k_{n}$ is the wave-vector mismatch, and $\tilde{\chi}%
_{n}\equiv 8\pi \omega _{n}^{2}/c^{2}\chi ^{(2)}$, $\chi ^{(2)}$
being an element of the quadratic susceptibility tensor. The extra
factors of $2$ in front of the coefficients $\tilde{\chi}_{1}$ and
$\tilde{\chi}_{3}$ in the first two equations reflects the fact
the equations may be derived from a
Lagrangian, and the factor $\sqrt{2}$ in front of all the terms containing $%
\tilde{\chi}_{2}$ is added by definition, to simplify subsequent rescalings.

Equations (\ref{startsys}) are normalized by measuring $x$ and $z$,
respectively, in units of the input beam size $r_{0}$ and diffraction length
$z_{d}=r_{0}^{2}k_{4}$ at the frequency $\omega _{1}+\omega _{2}$.
Introducing dimensionless fields
\begin{eqnarray}
E_{1} &=&(u_{1}/\sqrt{2}\gamma )\exp \left( i\beta _{1}z\right)
,E_{2}=(u_{2}/\sqrt{2}\gamma )\exp \left( i\beta _{2}z\right)
,E_{3}=(u_{3}/2\gamma )\exp \left( 2i\beta _{1}z-i\Delta k_{311}z\right) , \\
E_{4} &=&(u_{4}/\sqrt{2}\gamma )\exp \left( i(\beta _{1}+\beta
_{2})z-i\Delta k_{412}z\right) ,E_{5}=(u_{5}/2\gamma )\exp \left( 2i\beta
_{2}z-i\Delta k_{522}\right) ,
\end{eqnarray}
and $\gamma \equiv \tilde{\chi}_{1}r_{0}^{2}$, a normalized system of
equations is obtained:

\begin{eqnarray}
2\frac{k_{1}}{k_{4}}\left( i\frac{\partial u_{1}}{\partial z}-\beta
_{1}u_{1}\right) +\frac{\partial ^{2}u_{1}}{\partial x^{2}}+\chi
_{1}u_{1}^{\ast }u_{3}+\chi _{2}u_{2}^{\ast }u_{4} &=&0,  \nonumber \\
2\frac{k_{2}}{k_{4}}\left( i\frac{\partial u_{2}}{\partial z}-\beta
_{2}u_{2}\right) +\frac{\partial ^{2}u_{2}}{\partial x^{2}}+\chi
_{2}u_{1}^{\ast }u_{4}+\chi _{3}u_{2}^{\ast }u_{5} &=&0,  \nonumber \\
2\frac{k_{3}}{k_{4}}\left( i\frac{\partial u_{3}}{\partial z}-\left( 2\beta
_{1}-\Delta k_{311}\right) u_{3}\right) +\frac{\partial ^{2}u_{3}}{\partial
x^{2}}+\frac{1}{2}\chi _{1}u_{1}^{2} &=&0,  \label{middlesys} \\
2\frac{k_{4}}{k_{4}}\left( i\frac{\partial u_{4}}{\partial z}-\left( \beta
_{1}+\beta _{2}-\Delta k_{412}\right) u_{4}\right) +\frac{\partial ^{2}u_{4}%
}{\partial x^{2}}+\chi _{2}u_{1}u_{2} &=&0,  \nonumber \\
2\frac{k_{5}}{k_{4}}\left( i\frac{\partial u_{5}}{\partial z}-\left( 2\beta
_{2}-\Delta k_{522}\right) u_{5}\right) +\frac{\partial ^{2}u_{5}}{\partial
x^{2}}+\frac{1}{2}\chi _{3}u_{2}^{2} &=&0.  \nonumber
\end{eqnarray}
where $\chi _{n}\equiv \tilde{\chi}_{n}/\tilde{\chi}_{1}$ ($n=1,2,3$, i.e., $%
\chi _{1}\equiv 1$), $\beta _{1,2}$ being two phase-velocity shifts. To
reduce the number of parameters in Eqs. (\ref{middlesys}), the fields and
coordinates can be rescaled further. Defining $u_{n}=\beta _{1}U_{n}$, $%
z=Z/\beta _{1}$ , and $x=X/\sqrt{|\beta _{1}|}$, we obtain
\begin{eqnarray}
i\frac{\partial U_{1}}{\partial Z}+\frac{\partial ^{2}U_{1}}{\partial X^{2}}%
-\alpha _{1}U_{1}+\chi _{1}U_{1}^{\ast }U_{3}+\chi _{2}U_{2}^{\ast }U_{4}
&=&0,  \nonumber \\
i\frac{\partial U_{2}}{\partial Z}+\frac{\partial ^{2}U_{2}}{\partial X^{2}}%
-\alpha _{2}U_{2}+\chi _{2}U_{1}^{\ast }U_{4}+\chi _{3}U_{2}^{\ast }U_{5}
&=&0,  \nonumber \\
2i\frac{\partial U_{3}}{\partial Z}+\frac{\partial ^{2}U_{3}}{\partial X^{2}}%
-\alpha _{3}U_{3}+\frac{1}{2}\chi _{1}U_{1}^{2} &=&0,  \label{mainsys} \\
2i\frac{\partial U_{4}}{\partial Z}+\frac{\partial ^{2}U_{4}}{\partial X^{2}}%
-\alpha _{4}U_{4}+\chi _{2}U_{1}U_{2} &=&0,  \nonumber \\
2i\frac{\partial U_{5}}{\partial Z}+\frac{\partial ^{2}U_{5}}{\partial X^{2}}%
-\alpha _{5}U_{5}+\frac{1}{2}\chi _{3}U_{2}^{2} &=&0.  \nonumber
\end{eqnarray}
where $\alpha _{1}=1$, $\alpha _{2}=\beta _{2}/\beta _{1}$,
$\alpha _{3}=(4\beta _{1}-2\Delta k_{311})/\beta _{1}$, $\alpha
_{4}=(2\beta _{1}+2\beta _{2}-2\Delta k_{412})/\beta _{1}$, and
$\alpha _{5}=(4\beta _{2}-2\Delta k_{522})/\beta _{1}$. We assume
that $\left| k_{4}-k_{2}-k_{1}\right| \ll k_{n}$, and everywhere,
except for the phase-mismatch parameters introduced above, one may
set $k_{3}=k_{4}=k_{5}$, i.e., the three components of the
second-harmonic wave have similar wave numbers.

Equations (\ref{mainsys}) assume that the three parametric interactions
(``vertices''), which couple, respectively, the wave triplets (1,1,3),
(2,2,5), and (1,2,4), are nearly phase matched (as it was mentioned above, a
possible way of achieving this may be provided by the QPM technique based on
quasiperiodic structures \cite{Fradkin02}). It is straightforward to see
that, like the model describing the type-II SHG \cite
{Tran95,Malomed96,Buryak97}, Eqs. (\ref{mainsys}) have two Manley-Rowe
invariants, namely, the total power,

\begin{equation}
Q=\int_{-\infty }^{\infty }|U_{1}|^{2}+|U_{2}|^{2}+4\left(
|U_{3}|^{2}+|U_{4}|^{2}+|U_{5}|^{2}\right) dx,  \label{Q}
\end{equation}
and the power imbalance,
\begin{equation}
Q_{{\rm imb}}=\int_{-\infty }^{\infty }|U_{1}|^{2}-|U_{2}|^{2}+4\left(
|U_{3}|^{2}-|U_{5}|^{2}\right) dx.  \label{Qim}
\end{equation}

The present model does not include walkoff terms (group-velocity
mismatch). While a detailed discussion of the walkoff is beyond
the scope of this work, it is relevant to mention that QPM and
similar techniques, such as tandem structures \cite{tandem}, make
it possible to suppress the walkoff \cite {tandem,Stolzenberger}.
We also note that QPM can give rise to an effective cubic
nonlinearity \cite{Clausen}, which may compete with the underlying
quadratic interactions \cite{Bang,Johansen}. For this reason,
cubic terms should sometimes be added to a dynamical model, but
this is not an issue for immediate consideration in the present
context. As concerns the physical realization of the system,
estimates using typical values of the relevant physical parameters
in such quadratically nonlinear materials as lithium niobate and
KTP \cite{Fradkin02,Johansen} suggest that a necessary (quasi)
period of the QPM structure is $\ \sim 10$ $\mu $m, and the power
and transverse \ size of the soliton beam are expected to be
$\,\sim 20$ $\mu $m and $100$ mW, respectively.

If still more resonances are allowed, other essential wave components may
appear, for instance, those corresponding to the combinational frequencies $%
\omega _{12}\equiv 2\omega _{1}-\omega _{2}$ and $\omega _{21}\equiv 2\omega
_{2}-\omega _{1}$ (obviously, they belong to the fundamental-harmonic wave
set). Denoting the corresponding wave numbers as $k_{12}$ and $k_{21}$, we
see that these new components will indeed be essential if the wave-number
triplets (12,2,3) and/or (21,1,5) are nearly phase-matched. The accordingly
modified system will include seven components (four pertaining to the
fundamental harmonic, and three to the second harmonic) and five vertices.
However, such a situation seems much more exotic (five simultaneous
resonances) than the more generic possibility of three simultaneous
resonances underlying the model considered in the present work.

\section{Stationary soliton solutions}\label{three}

Particular exact solutions of Eqs. (\ref{mainsys}) for stationary solitons
can be sought for in the form
\begin{equation}
U_{n}=A_{n}{\rm sech}^{2}\left( \lambda X\right) ,\,n=1,...,5,
\label{ansatz}
\end{equation}
where $A_{n}$ are amplitudes, and $\lambda $ is the inverse width of the
soliton. Inserting this (\ref{ansatz}) into Eqs. (\ref{mainsys}), a solution
can be obtained if $\alpha _{1}=\alpha _{2}=\alpha _{3}=\alpha _{4}=\alpha
_{5}\equiv \alpha $ and $\lambda =\sqrt{\alpha }/2$. The amplitudes are
found to be
\begin{eqnarray}
A_{1} &=&\pm \sqrt{\xi }A_{2},\,A_{2}=\pm \frac{6\lambda ^{2}}{\chi
_{2}^{2}\xi +\chi _{3}^{2}/2},\,A_{3}=\frac{\chi _{1}A_{1}^{2}}{12\lambda
^{2}},  \nonumber \\
A_{4} &=&\frac{\chi _{2}A_{1}A_{2}}{6\lambda ^{2}},\,A_{5}=\frac{\chi
_{3}A_{2}^{2}}{6\lambda ^{2}},  \label{general_sol}
\end{eqnarray}
where $\xi \equiv (\chi _{3}^{2}/2-\chi _{2}^{2})/(\chi _{1}^{2}/2-\chi
_{2}^{2})$.

In addition to the exact solution based on Eqs. (\ref{ansatz}) and (\ref
{general_sol}), other particular solutions can be found setting $U_{2,4,5}=0$
or, alternatively, $U_{1,3,4}=0$. Equations (\ref{mainsys}) then reduce to
the well-known type-I SHG model \cite{Stegeman,Buryak95,Etrich01},
\begin{eqnarray}
i\frac{\partial V}{\partial Z}+\frac{\partial ^{2}V}{\partial X^{2}}%
-V+V^{\ast }W &=&0,  \label{FF} \\
2i\frac{\partial W}{\partial Z}+\frac{\partial ^{2}W}{\partial X^{2}}-\rho W+%
\frac{1}{2}V^{2} &=&0,  \label{SH}
\end{eqnarray}
where $V$ and $W$ are the fields at the fundamental and second harmonics. In
this case, there is a well-known special exact solution corresponding to $%
\rho =1$ in Eq. (\ref{SH}) \cite{Karamzin74},
\begin{equation}
V=\left( 3/\sqrt{2}\right) {\rm sech}^{2}\left( X/2\right) ,\,W=\left(
3/2\right) {\rm sech}^{2}\left( X/2\right) .  \nonumber
\end{equation}
When $\rho \neq 1$, a family of stationary soliton solutions to Eqs. (\ref
{FF}) and (\ref{SH}) can be found numerically \cite{Buryak95} (or
approximated analytically by means of the variational method \cite{Vika}).
We will refer to the general solution for the case when $U_{1}=V$, $U_{3}=W$%
, and $U_{2,4,5}=0$ as an ``A'' type soliton, while the opposite ``B'' type
is defined as the one with $U_{2}=V$, $U_{5}=W$, and $U_{1,3,4}=0$. Both
these types will also be called ``simple'' solitons.

\begin{figure}[tbp]
\includegraphics[scale=0.4]{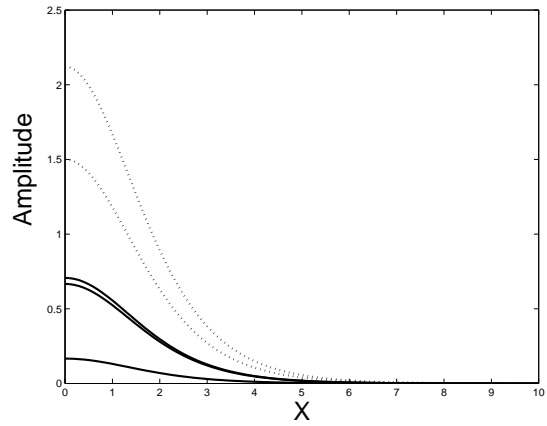}
\caption{Profiles of a polychromatic soliton (solid lines) and a type ``A''
soliton (dotted). Common values of the parameters for these solutions are $%
\protect\alpha_{1,...,5}=1$, $\protect\chi_1=1$, $\protect\chi_2=2$, and $%
\protect\chi_3=1$.}
\label{compare}
\end{figure}

\begin{figure}[tbp]
\includegraphics[scale=0.4]{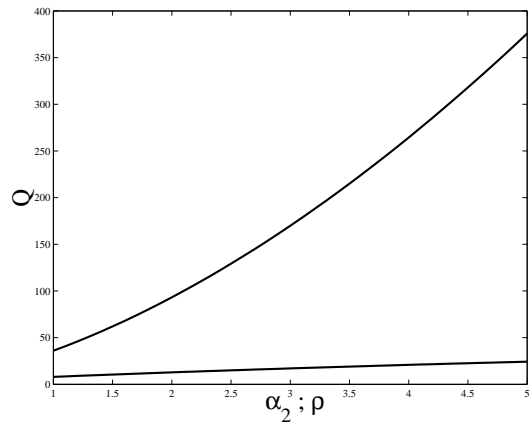}
\caption{The power invariant $Q$ for both the ``A''-type (upper line, vs. $%
\protect\rho$) and full polychromatic (lower line, vs. $\protect\alpha_2$)
solitons. Values of the parameters are the same as in Fig. \ref{compare}.}
\label{qbeta}
\end{figure}

General solutions for the polychromatic (five-wave) soliton can be found
from the stationary version (the one with $\partial U_{n}/\partial Z=0$) of
Eqs. (\ref{mainsys}) by means of the standard numerical methods for
two-point boundary-value problems. In Figs. \ref{compare} and \ref{qbeta},
comparison is made between the general five-wave solitons and the particular
solutions generated by Eqs. (\ref{FF}) and (\ref{SH}) at equal values of all
the parameters. In Fig. \ref{compare}, one can see that the A soliton is
much larger in amplitude, while its width is not widely different from that
of the polychromatic one. As a consequence of this, the power $Q$ [see Eq. (%
\ref{Q})] of the A-type soliton shown in Fig. \ref{compare} is $Q=36$, while
for PCS in the same figure, $Q=8$ [the other invariant is $Q_{{\rm imb}}=0$
in the case considered, see Eq. (\ref{Qim})].

This drastic difference in the powers can be understood: in the
case of the full PCS, one has two nonlinear terms in the first two
equations (\ref {mainsys}), rather than one term in the case of
the simple solitons; therefore, the amplitude necessary to
compensate the spreading out of the beam due to the diffraction
term is, roughly, twice as small in comparison with the simple
solitons, or, eventually, the power is $\,\sim 4$ times as small.
To check whether the power of the PCS is indeed essentially lower
than that of simple solitons in the general case, in Fig.
\ref{qbeta} we show a typical example of the change of the two
powers with the variation of $\alpha _{2}$ or, equivalently, $\rho
$ in Eq. (\ref{SH}). As is seen, PCS indeed persistently maintains
a lower power $Q$ than its A-type counterpart. On varying the
different parameters $\alpha_n,\chi_n$ the value of $Q$ of the
polychromatic soliton may be increased (or decreased) in value
from that portrayed in Fig. \ref{qbeta} but it only exceeds that
of the simple solitons when $\chi_2<0.7$. In the limit of
$\chi_2\rightarrow 0$ Eqs. (\ref{mainsys}) decouple and the
polychromatic soliton tends to a ``double" simple soliton i.e.
both ``A" and ``B" type solitons existing in one envelope. The
double simple soliton by definition has twice the power of single
simple soliton. Provided the nonlinearity coefficient $\chi_2$ is
large enough ($\chi_2\geq 0.7$) it is possible to conclude that
PCSs in a quadratically nonlinear medium may be produced from
waves with different frequencies at a much {\em lower} net input
power than the ordinary SHG\ solitons, i.e., it may be essentially
easier to generate PCSs in the experiment and use them in
potential applications. Of course, these results are meaningful
provided that these solitons are stable.

\section{Stability and interactions of the solitons}\label{four}

\subsection{The stability of the polychromatic solitons}

To test the stability of PCSs, we solved the full system of Eqs. (\ref
{mainsys}), employing the numerical split-step fast-Fourier-transform
method. The simulations were performed with a computational grid of 2048
points, the transverse and propagation step-sizes being, respectively, $%
h_{x}\approx 0.02$ and $h_{z}=0.01$. The integration domain had the
transverse size $100$ ($-50<x<+50$), which is by far larger than any $x$
size relevant to the solitons, see Figs. 4, 5, and 9 below. Absorbing layers
were placed at the edges of the computational domain to prevent reentering
of radiation. Specially monitoring interaction of the radiation waves with
the absorbers, we have verified that, in all the cases considered, no
reflection took place indeed.

Evolution of initial configurations close to the stationary solutions was
simulated for a variety of parameters. To impose initial perturbations,
values of the initial amplitude and width of the pulses were altered against
the exact stationary solutions. From the results of the simulations, we have
concluded that PCSs survive, clearly remaining stable, as long as the
simulations were run, the maximum simulation length being 300 diffraction
lengths of the soliton.

\begin{figure}[tbp]
\includegraphics[scale=0.4]{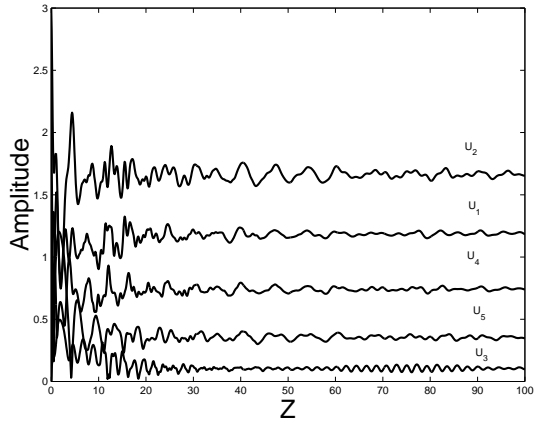}
\caption{The generation of a polychromatic soliton from the Gaussian input $%
U_1=2e^{-x^2/4}$ and $U_2=3e^{-x^2/4}$. After an initial period of strong
relaxation, a weakly oscillating soliton is produced. The evolution of the
amplitudes of the $U_n$ components vs. $Z$ is plotted. The parameters used
in this simulation were $\protect\alpha_1=\protect\alpha_2=1$, $\protect%
\alpha_3= \protect\alpha_4= \protect\alpha_5=4$, $\protect\chi_1=\protect\chi%
_3 =1$, and $\protect\chi _2=2$. The power invariants are $Q=13\protect\sqrt{%
2\protect\pi}$ and $Q_{imb}=-5\protect\sqrt{2\protect\pi}$.}
\label{maxpt}
\end{figure}

To further test robustness of PCSs, we ran numerical experiments
in which the solitons were successfully generated from initial
Gaussian pulses, launched in the pump fields $U_{1}$ and $U_{2}$
(with no initial second-harmonic components, which corresponds to
the generation of SHG solitons in real experiments
\cite{Stegeman_experiment}). The results demonstrate that PCSs not
only are stable against small perturbations, but also play the
role of strong attractors in the system. An example of PCS
generation from the Gaussian beams is illustrated by Fig.
\ref{maxpt}. Typically, there is a period of strong relaxation of
the beams, where the amplitudes fluctuate and excess energy is
radiated away. The second-harmonic components $U_{3,4,5}$ are
generated, and the soliton arranges itself to a quasi-stationary
form, which then propagates, in a stable fashion, over a distance
in excess of 100 diffraction lengths (as long as the simulations
were run). As a result of many runs of systematic simulations,
PCSs have been found to be attractors in a broad range in the
system's parameter space. We concentrated mainly on the parameter
space $\alpha_n = 1..5$ and $\chi_n = 1..2$ where the PCS
definitely had a lower value $Q$ than the simple solitons. In
connection to this, it is relevant to note that the
above-mentioned families of the simple A and B solitons are also
stable in the ordinary SHG model, except for a small region of
their existence domain \cite{Buryak95,Etrich01}.

\subsection{Collisions between orthogonal simple solitons}

An interesting possibility is to consider collisions of the mutually
orthogonal simple solitons of the above-mentioned A and B types. The
overlapping between the colliding solitons will give rise to the generation
of the field $U_{4}$, which is absent in both A and B solitons in their pure
form, and the issue is how the generation of this field will affect the
interaction between the solitons. Equations (\ref{SH}) have the property of
the Galilean invariance, so ``moving'' solitons (in fact, the solitary beams
propagating at an angle in the planar waveguide) can be constructed by the
transformation
\begin{eqnarray}
V(X,Z) &=&V_{0}\left( X-CZ\right) \exp \left( iCX/2-iC^{2}Z/4\right) \,,
\label{transform} \\
W(X,Z) &=&W_{0}\left( X-CZ\right) \exp \left( iCX-iC^{2}Z/2\right) \,,
\nonumber
\end{eqnarray}
where $C$ is the ``velocity'' (actually, a slope) of the ``moving'' soliton.

We collided the A and B solitons with an initial separation between their
centers $x_{0}=20$, varying their velocities $\pm C$. A representative set
of values of the parameters for which the results are displayed below are $%
\alpha _{1,...,5}=1$, $\chi _{1}=\chi _{3}=1$, and $\chi _{2}=2$; many
simulations run for other values gave quite similar results.

\begin{figure}[tbp]
\includegraphics[scale=0.4]{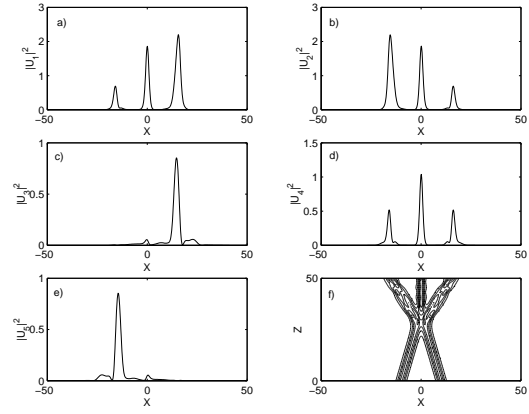}
\caption{The collision between simple solitons of the mutually orthogonal A
and B types, with the initial velocities $\pm 0.3$ and zero phase
difference. The parameters are $\protect\alpha_n=1$, $\protect\chi_1=\protect%
\chi_3=1$, $\protect\chi_2=2$, and $C=0.3$. The field components are
displayed in the panels: $U_1$ (a), $U_2$ (b), $U_3$ (c), $U_4$ (d), and $U_5
$ (e). The panel (f) shows a combined contour plot of the $U_1$ and $U_2$
amplitudes. Note the mirror symmetry of the profiles of the components $U_1$
and $U_2$, which is a consequence of the fact that the imbalance invariant $%
Q_{imb}$ is zero in this case. It is obvious that the simple solitons become
polychromatic after the collision, and the third polychromatic soliton with
the zero velocity is generated.}
\label{asymm3}
\end{figure}

\begin{figure}[tbp]
\includegraphics[scale=0.4]{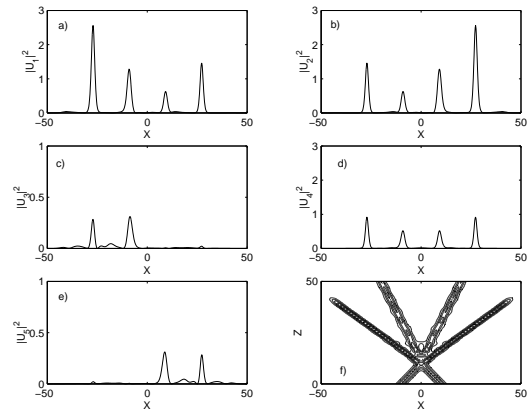}
\caption{The same as in Fig. \ref{asymm3}, except that the collision
velocities are larger, $\pm 1.0$. A zero-velocity soliton is no longer
generated after the collision. Instead, four solitons are observed in the
post-collision state. In the panel (f) the outer beams terminate prematurely
because they hit the absorbing sponges used in the numerical scheme.}
\label{asymm100}
\end{figure}

In the case of moderate collision velocities (and zero phase difference
between the solitons), the generation of the field $U_{4}$ in the course of
the collision gives rise to a third {\em polychromatic} soliton with the
zero velocity. Trajectories of the initial A and B solitons alter, and they
appear from the collisions as PCSs too, i.e., the collision adds to them
components which were initially absent. The three post-collision solitons
have roughly equal powers, with a mirror symmetry in the power distribution
amongst individual components, see Fig. \ref{asymm3}. At higher collision
velocities (again, for the zero phase difference), {\em four} PCSs are
generated by the collision, all having a nonzero velocity (see Fig. \ref
{asymm100}). As the collision velocity increases, less and less interaction
takes place, until the initial solitons pass through one another unchanged
(elastically). In all these cases, 90$\%$ of the initial power is typically
converted into the resulting PCSs, i.e., radiation loss is not a dominant
actor in the collision dynamics.

\begin{figure}[tbp]
\includegraphics[scale=0.4]{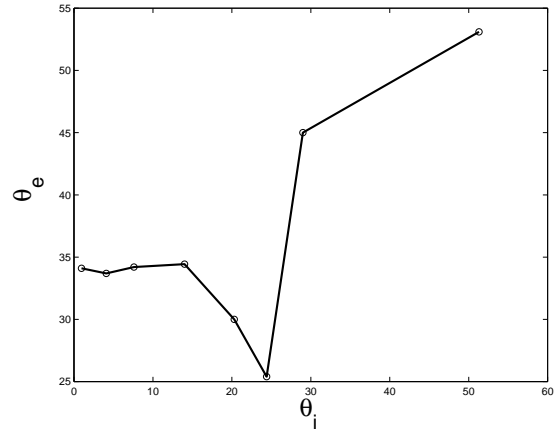}
\caption{The exit angle (in degrees) $\protect\theta_{{\rm e}}$ for the
outer solitons, as found from the simulations of the collision, vs. the
incidence angle $\protect\theta_{{\rm i}}$. The angles are between the
solitons' trajectories and the $z$ axis in the $(x,z)$ plane, so that an
increase in the incidence angle corresponds to an increase in the collision
velocity. Naturally, the collision becomes less inelastic as $\protect\theta%
_{{\rm i}}$ increases, and eventually the exit angle becomes nearly equal to
the incidence one.}
\label{angles}
\end{figure}

\begin{figure}[tbp]
\begin{center}
\includegraphics[scale=0.4]{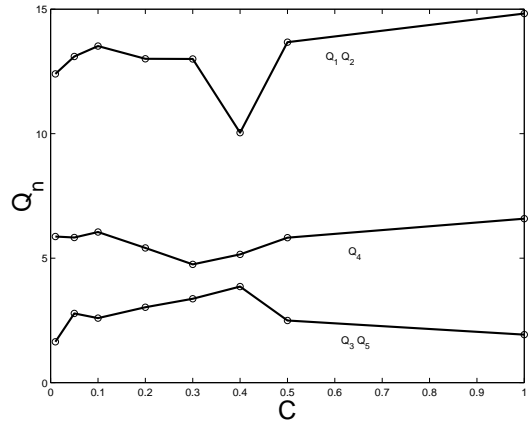}
\end{center}
\caption{Distribution of the powers $Q_n=\int_{-\infty}^\infty |U_n|^2 dx$
between the components of the former A or B soliton after the collision, as
a function of the soliton velocity before the collision. At higher
velocities, the collision gives rise to 4 (rather than 3) solitons, which
take a part of the energy.}
\label{distro}
\end{figure}

Results produced by many runs of the simulations for the collision of the
in-phase (zero-phase-difference) solitons are summarized in Figs. \ref
{angles} and \ref{distro}. The constant initial separation $x_{0}$ between
the solitons in the simulations means that the increasing velocity
corresponds to an increasing incidence angle $\theta _{{\rm i}}$. As may be
naturally expected, the collision alters the trajectories of the solitons
most in the case of low velocities. At higher velocities, the outer solitons
keep essentially the same velocity after the collision as they had before
it, while the two additionally generated inner solitons are significantly
slower.

The most significant inelastic effects occur when the solitons collide with
the velocities $\pm 0.4$. In this case, the exit trajectory is altered the
most (see Fig. \ref{angles}), and, as per Fig. \ref{distro}, the largest
part of the net power is transferred into the newly generated harmonic
components of the outer solitons.

\begin{figure}[tbp]
\includegraphics[scale=0.4]{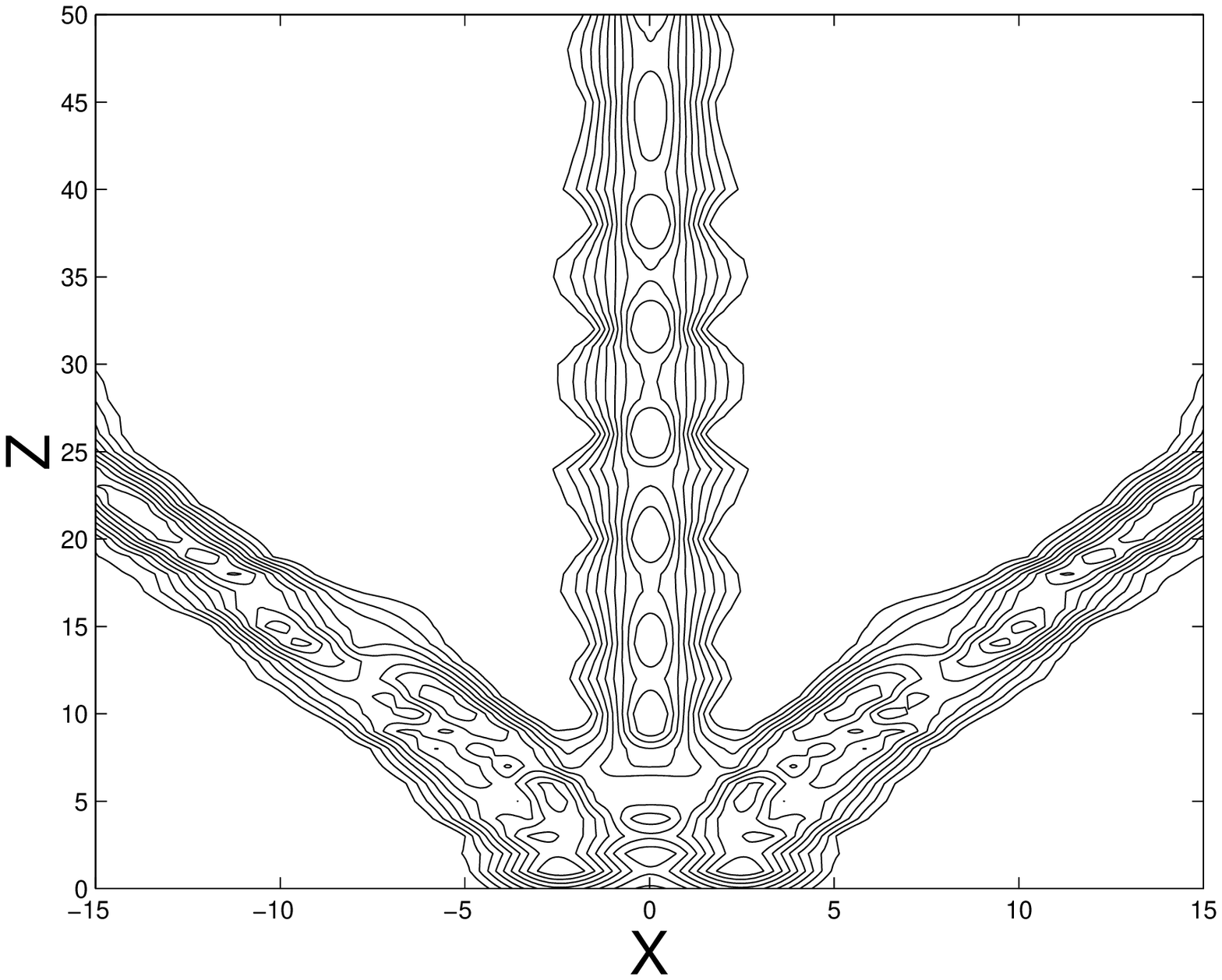}
\caption{The combined amplitude contour plot of the pump waves $U_1$ and $%
U_2 $ in the case when the initial velocities of the simple solitons ``A''
and ``B'' are zero.}
\label{FFzerovel}
\end{figure}

The interaction between the A and B solitons produces nontrivial results
also in the case when they have zero initial velocities, but their tails
overlap at the initial position. For the initial separation $x_{0}=5$, the
result of the interaction with $C=0$ is displayed in Fig. \ref{FFzerovel},
in the form of the distribution of the fields $U_{1}$ and $U_{2}$ produced
by the interaction. Again, three PCSs appear, the central one with the zero
velocity, and two outer solitons with nonzero velocities. In the course of
the evolution, a phase difference develops across the component fields. In
fact, it is this phase difference which repulses the outer solitons, lending
them a nonzero velocity.

It is well known from theoretical \cite{Stegeman_theory} and
experimental \cite{Stegeman_experiment} studies of collisions
between SHG\ solitons of the usual type that the result strongly
depends on their relative phase at the collision point (provided
that the colliding solitons have nearly identical amplitudes):
they attract each other and therefore interact strongly in the
case of the zero phase difference, and repel each other if the
phase difference between the fundamental-harmonic components is
$\pi$. However, in the present system the effect of the phase
difference on collisions between the simple solitons of the
orthogonal types is much weaker. Simulations performed with
various phase differences (including $\pi/2$) between the two
orthogonal simple solitons does not display any change in the
post-collision dynamics of the beams when compared to the zero
phase difference case. Also it may be seen from the structure of
the underlying system (\ref{mainsys}) that phase differences
between the ``A" and ``B" solitons will have little effect.
Indeed, these
equations are exactly invariant against the transformation $%
u_{1}\rightarrow u_{1}\exp \left( i\phi_1 \right) $,
$u_{3}\rightarrow u_{3}\exp \left( 2i\phi_1 \right) $,
$u_{2}\rightarrow u_{2}\exp \left( i\phi_2 \right) $,
$u_{5}\rightarrow u_{5}\exp \left( 2i\phi_2 \right) $ and
$u_{4}\rightarrow u_{4}\exp \left( i\phi_1+i \phi_2 \right) $ with
an arbitrary phase shift $\phi_{1,2} $.

The interactions between the simple solitons in this system may find
application as a basis for an all-optical logic gate. Indeed, consider two
solitons of the ``A'' and/or ``B'' types, propagating in such a way that
they will collide. If the solitons belong to the same type (i.e., the
configuration is AA or BB), then they attract or repel one another,
depending on their relative phase \cite{Stegeman_experiment,Stegeman_theory}%
, or, if the relative velocity is high enough, they simply pass through one
another. But if they belong to the opposite types (an AB configuration), at
least three PCSs are produced by the collision even at high velocities. This
is a behavior which is expected from an exclusive OR gate, alias XOR. The
actual outcome can be easily established by checking the $U_{4}$ content in
the output (by means of an appropriate optical filter). An advantage of such
a design of the XOR gate is that any output beam will be a stationary
soliton (even in the case of a strongly inelastic collision), which makes it
convenient for further manipulations (cascadability).

\section{Conclusion}\label{five}

We have introduced the simplest five-component model of polychromatic
solitons (PCSs) in a quadratically nonlinear optical medium. The existence
and stability of both polychromatic and two types of simple solitons have
been demonstrated. An essential peculiarity of PCSs is that their power is
much smaller, at the same values of the control parameters, than the power
of the usual two-component (simple) solitons. We have also performed
systematic simulations of collisions between mutually orthogonal simple
solitons, concluding that the collisions are strongly inelastic (including
the interaction between two solitons with the zero initial velocity), giving
rise to transformation of the simple solitons into polychromatic ones, and
generation of one or two additional PCSs. A value of the relative velocity
at which the inelastic effects are strongest has been found. We have also
shown that the collision may serve as a basis to design a simple all-optical
logic gate of the XOR type.

\begin{acknowledgments}
We appreciate valuable discussions with A.V. Buryak. This work was
supported, in a part, by the US-Israel Binational Science Foundation
under the grant No. 1999459, and by a matching grant from the
Tel Aviv University.
\end{acknowledgments}

\begin{thebibliography}{99}
\bibitem{Franken61}  P.A. Franken, A.E. Hill, C.W. Peters, and G. Weinreich, %
\prl{\bf 7}, 118 (1961).

\bibitem{Armstrong62}  J.A. Armstrong, N. Bloembergen, J. Ducuing, and P.S.
Pershan, Phys. Rev. {\bf 127}, 1918 (1962).

\bibitem{Etrich01}  C. Etrich, F. Lederer, B.A. Malomed, T. Peschel, and U.
Peschel, Progr. Optics {\bf 41}, 483 (2000).

\bibitem{Torruellas95}  W.E. Torruellas, Z. Wang, D.J. Hagan, E.W. Van
Stryland, G.I. Stegeman, L. Torner, and C.R. Menyuk, \prl
{\bf 74}, 5036 (1995).

\bibitem{Schiek96}  R. Schiek, Y. Baek, and G.I Stegeman, \pre
{\bf 53}, 1138 (1996).

\bibitem{Stegeman}  C.R. Menyuk, R. Schiek, and L. Torner, J. Opt. Soc. Am.
B {\bf 11}, 2434 (1994).

\bibitem{Buryak95}  A.V. Buryak and Y.S. Kivshar, \pl A {\bf 197}, 407
(1995).

\bibitem{Tran95}  H.T. Tran, \oc{\bf 118}, 581 (1995).

\bibitem{Malomed96}  B.A. Malomed, D. Anderson, and M. Lisak, \oc{\bf 126}
251 (1996).

\bibitem{Buryak97}  A.V. Buryak, Y.S. Kivshar, and S. Trillo,
\josab
{\bf 14}, 3110 (1997).

\bibitem{JenaTypeII}  U. Peschel, C. Etrich, F. Lederer, and B.A. Malomed, %
\pre{\bf 55}, 7704 (1997).

\bibitem{Boardman}  A.D. Boardman, P. Bontemps, and K. Xie, Opt. Quant.
Electron. {\bf 30}, 891 (1998).

\bibitem{Fradkin99}  K. Fradkin-Kashi and A. Arie, IEEE J. Quantum Electron.
{\bf 35}, 1649 (1999).

\bibitem{Fradkin02}  K. Fradkin-Kashi, A. Arie, P. Urenski, and G. Rosenman, %
\prl{\bf 88}, 023903 (2002).

\bibitem{Johansen}  S.K. Johansen, S. Carrasco, L. Torner, and O. Bang, Opt.
Commun. {\bf 203}, 393 (2002).

\bibitem{Kivshar99a}  Y.S. Kivshar, T.J. Alexander, and S. Saltiel,
\ol
{\bf 24}, 759 (1999).

\bibitem{Kivshar99b}  Y.S. Kivshar, A.A. Sukhorukov, and S. M. Saltiel,
\pre
{\bf 60}, R5056 (1999).

\bibitem{Towers00}  I. Towers, A.V. Buryak, R.A. Sammut, and B.A. Malomed, %
\josab{\bf 17}, 2018 (2000).

\bibitem{Sukhorukov01}  A.A. Sukhorukov, T.J. Alexander, Y.S. Kivshar, and
S.M. Saltiel, \pl A {\bf 281}, 34 (2001).

\bibitem{Agrawal}  G.P. Agrawal. {\it Nonlinear Fiber Optics} (Academic
Press: San Diego, 1995).

\bibitem{tandem}  L. Torner, IEEE Photonics Techn. Lett. {\bf 11}, 1268
(1999).

\bibitem{Stolzenberger}  R. Stolzenberger, Lasers \& Optronics {\bf 19}, 17 (2000).

\bibitem{Clausen} C. B. Clausen, O. Bang and Y. S. Kivshar, \prl
{\bf 78}, 4749 (1997).

\bibitem{Bang}  O. Bang, C.B. Clausen, P.L. Christiansen, and L. Torner,
Opt. Lett. {\bf 24}, 1413 (1999); O. Bang, T.W. Graversen, and J.F. Corney,
Opt. Lett. {\bf 26}, 1007 (2001).

\bibitem{Karamzin74}  Y.N. Karamzin and A.P. Sukhorukov, JETP Lett. {\bf 20}%
, 339, (1974).

\bibitem{Vika}  V. Steblina, Yu.S. Kivshar, M. Lisak, and B.A. Malomed,
\oc
{\bf 118} 345, (1995).

\bibitem{Stegeman_experiment}  R. Schiek, Y. Baek, G. Stegeman, and W.
Sohler, Opt. Quant. Electr. {\bf 30}, 861 (1998).

\bibitem{Stegeman_theory}  D.M. Baboiu and G.I. Stegeman, Opt. Quant.
Electr. {\bf 30}, 849 (1998).
\end{thebibliography}
\end{document}